# Testing Self-Organized Criticality by Induced Seismicity

Jean-Robert Grasso

Observatoire de Grenoble, LGIT/IRIGM Grenoble, France

Didier Sornette

Institute of Geophysics and Planetary Physics and Department of Earth & Space ScienceUniversity of California, Los Angeles
1

**Abstract.** We examine the hypothesis proposed in recent years by several authors that the crust is in a self-organized critical (SOC) state. This hypothesis has been suggested on the basis of the observation of power law distributions, such as the Gutenberg-Richter law for earthquakes and the fault length distribution, and of the fractal geometry of sets of earthquake epicenters and of fault patterns. These self-similar properties are shared by simplified models of the crust exhibiting a spontaneous organization toward a critical point characterized by similar scale-invariant properties. The term "critical" is here used in the sense of phase transitions such as the Curie point in magnetism. The usefulness of a hypothesis is measured by its predictive and explanatory power outside the range of observations that have helped defined it. We thus explore how the SOC concept can help in understanding the observed earthquake clustering on relatively narrow fault domains and the phenomenon of induced seismicity. We review the major reported cases of induced seismicity in various parts of the world and find that both pore pressure changes ($\pm\Delta p$) and mass transfers ($\pm\Delta m$) leading to incremental deviatoric stresses of <1 MPa are sufficient to trigger seismic instabilities in the uppermost crust with magnitude ranging up to 7.0 in otherwise historically aseismic areas. Once triggered, stress variations of at least 1 order of magnitude less but still larger than the ~0.01 MPa tidal stress are enough to sustain seismic activity. We argue that these observations are in accord with the SOC hypothesis as they show that a significant fraction of the crust is not far from instability and can thus be made unstable by minute perturbations. This property is shared by simplified models of SOC. Not all perturbations, however, trigger seismic activity; this is also compatible with the SOC hypothesis which embodies naturally the existence of large heterogeneities in the stress field. The induced seismicity is found to obey generally the Gutenberg-Richter law up to a magnitude cutoff which correlates well with the width of the local seismogenic bed, ranging in size from that of mine pillars for mining-induced seismicity to the thickness of brittle sedimentary beds in the vicinity of dams or depleted hydrocarbon reservoirs. In conclusion, the properties of induced seismicity and their rationalization in terms of the SOC concept provide further evidence that potential seismic hazards extend over a much larger area than that where earthquakes are frequent.

## 1. Introduction

Seismicity is characterized by an extraordinary rich phenomenology and variability which makes the development of a coherent explanatory and predictive framework very difficult. In the late 1980s, the concept of self-organized criticality (SOC) was put forward as a possible candidate [*Bak et al.*, 1987; *Sornette and Sornette*, 1989; *Bak and Tang*, 1989; *Ito and Matsuzaki*, 1990; *Sornette et al.*, 1990; *Scholz*, 1991, *Sornette*, 1991]. Apart from the rationalization that it provides for the Gutenberg-Richter law for earthquakes, for the power law fault length distribution, and for the fractal geometry of sets of earthquake epicenters and fault patterns, it has not been really exploited to advance our understanding of the crust organization and the very rich and subtle properties found in tectonics and seismology.

We shall first briefly review the basic idea of SOC and then discuss some of the expected properties of the SOC state that have been only slightly considered previously in the seismological context. Our goal is to explore the usefulness of the SOC hypothesis by making explicit novel properties and predictions that have not been used to define it and by testing them in the field. The two aspects that we address specifically are the problems of spatial localization of seismic activity and of induced seismicity.

Most of the active faults in the world are concentrated along plate boundaries, and slip on them results from the relative movements between essentially rigid plates. Stress accumulates continuously as one plate moves past another until an earthquake occurs. Yet earthquakes also occur within the generally aseismic rigid plates, and little is known about the processes causing intraplate seismic instabilities in continental setting nor, for that matter, in oceanic setting or even in plate boundaries. The mechanisms for seismogenic failures well away from plate boundaries are generally unclear, despite numerous studies [see, e.g., *Byerlee*, 1967; *Nur,* 1972; *Costain et al.*,1987 *Sibson*, 1989] that rely, by indirect evidence, on changes in fluid pressure and movements of fluids. A general strategy that has been extraordinarily successful in the physical sciences is to study a system by monitoring its response to external perturbations. In this sense, earthquakes induced by human activity provide a unique opportunity to apply this strategy at the geological scale for a better understanding of the physics of the crust. When the mechanics of earthquakes induced by human activity is known, an estimation of the perturbation of the stress field that drives faulting is possible in terms of driving shear stress. Previous works related seismic stress parameters (stress drops from induced earthquakes) to the relevant in situ stress and rock properties [*Wyss and Molnar*, 1972; *McGarr et al.*, 1979; *Fletcher,* 1982; *Simpson*, 1986]. In situ stress measurements in areas of active faulting and induced seismicity suggest that the preexisting state of stress in the crust is essentially in frictional equilibrium with the frictional strength of highly faulted rock in some part of the crust so that the small stress (or strength) perturbations associated with reservoir impoundment of fluid injection or withdrawal can trigger failure at these sites [see, e.g., *Zoback and Healy,* 1984].

We will review the major reported cases of induced seismicity, including reservoir-induced seismicity and seismicity triggered by subsurface fluid manipulations. A special emphasis is put on so-called "stable" tectonic settings, i.e. areas characterized by a deformation rate slower than that observed in the so-called "active" areas where most of the seismicity takes place. The source mechanisms of triggered earthquakes are used to quantify the local critical stress thresholds. Observations reported in Zambia [*Gough and Gough,* 1970], in SE China and in SW India [*Gupta and Rastogi*, 1976], in western Uzbekistan [*Simpson and Leith*, 1985; *Pearce*, 1987; *Bowers and Pearce*, 1995], in northern Holland, in Norway, and in SW France [*Grasso*, 1993] provide examples of large induced seismicity rates in regions with low historical seismicity rates, due to the vicinity of water reservoir impoundments or oil and gas extraction.

During the exposition and in the synthesis of these observations, we will outline how these observations and their interpretation can be compared to the predictions offered by the SOC hypothesis. To help the reader follow the argument, it is useful to summarize our main conclusion. The proposed "critical" nature of the state of the crust is characterized by the property of a large susceptibility, i.e. small external perturbations may lead to large responses in some parts of the system. The seismicity induced by weak relative stress changes can thus be rationalized in this framework. However, not all perturbations lead to seismic instabilities. This common observation is shown to also be in agreement with the prediction of large fluctuations in the stress field associated with the critical state. This stress heterogeneity is related to the seismic clustering and fault localization, introducing local screening and enhancement effects. Recent simplified models of the crust organization allow one to understand how a system can self-organize to a global critical state characterized by various power laws and a fractal geometry on one hand and exhibit localization on the other hand. This localization is seen to belong to the overall self-organization mechanism. This allows one to understand why some perturbations do not seem to destabilize the crust, in contradiction with the widespread (incorrect) idea that self-organized criticality implies that the crust is everywhere on the verge of rupturing.

## 2. Self-Organized Criticality

### 2.1. Definition of Self-Organized Criticality

Let us start with a rather abstract and general description of SOC, and then we will describe its specialization to the crust problem. In the literature, the term SOC has not always been used with the same meaning, and we think that in fact it has been often misused. To our point of view, for this concept to have some usefulness, it must be rather specific and refer to a well-defined situation based on physical mechanisms rather than on observations. In other words, a system is not a self-organized criticality solely because it exhibits a power law distribution of event sizes (see *Sethna et al.*

[1993], *Sornette* [1994a,b], and *Percovic et al.* [1995] for counterexamples). Roughly speaking, SOC refers to the spontaneous organization of a system driven from outside in a dynamical statistical stationary state, which is characterized by self-similar distributions of event sizes and fractal geometrical properties.

We thus restrict the term SOC to the class of phenomena occurring in continuously driven out-of-equilibrium systems made of many interactive components, which possess the following fundamental properties: (1) a highly nonlinear behavior, namely, essentially a threshold response, (2) a very slow driving rate, (3) a globally stationary regime, characterized by stationary statistical properties, and (4) power distributions of event sizes and fractal geometrical properties. The crust obeys these four conditions [*Sornette*, 1991]:

1. The threshold response can be associated with the stick-slip instability of solid friction or to a rupture threshold thought to characterize the behavior of a fault upon increasing applied stress.

2. The slow driving rate is that of the slow tectonic deformations thought to be exerted at the borders of a given tectonic plate by the neighboring plates and at its base by the underlying lower crust and mantle. The large separation of timescales between the driving tectonic velocity (~cm/yr) and the velocity of slip (~m/s) makes the crust problem maybe the best natural example of self-organized criticality, in this sense. It is important to realize that these two ingredients must come together. Let us imagine a system composed of elements with threshold dynamics, but which is driven at a finite rate compared to the typical time scale of its response. In the earthquake problem, this would correspond to imagining plates moving at a velocity which is not much smaller than the rupture front velocity of brittle failure. In this case, the crust would rupture incessantly, with earthquakes which could not be separated and which would create an average rapid flow deformation of the crust. The frequency-size distribution would disappears and the system would not be in a SOC state. Artificial block-spring models of rapidly driven seismicity can be used to illustrate this idea [*Burridge and Knopoff*, 1967; *Carlson et al.*, 1991]. Alternatively, consider a system driven very slowly, with elements which do not possess a threshold but only a strong, continuous, nonlinear response. Due to the very slow driving, at a given instant, only a small increment is applied to the system, which responds adiabatically without any interesting behavior such as earthquakes with brittle rupture ("slow" earthquake regime? [*Linde et al.*, 1996]). It is thus the existence of the threshold that enables the system to accumulate and store the slowly increasing stress until the instability is reached and an earthquake is triggered. In turn, the slow tectonic driving allows for a response which is decoupled from the driving itself and which reflects the critical organization of the crust.

3. The stationarity condition ensures that the system is not in a transient phase and distinguishes the long-term organization of faulting in the crust from, for instance, irreversible rupture of a sample in the laboratory.

4. The power laws and fractal properties reflect the notion of scale invariance; namely, measurements at one scale are related to measurements at another scale by a normalization involving a power of the ratio of the two scales. These properties are important and interesting because they characterize systems with many relevant scales and long-range interactions as probably exist in the crust.

A remarkable feature of the SOC state is in the way it emerges. The spatial correlations between different parts of the system do not appear to be a result of a progressive diffusion from a nucleus but a result from the repetitive action of rupture cascades (would be earthquakes). In other words, within the SOC hypothesis, different portions of the crust become correlated at long distances by the action of earthquakes which "transport" the stress field fluctuations in the different parts of the crust many times back and forth to finally organize the system. This physical picture is substantiated by various numerical and analytical studies of simplified models of the crust [*Bak and Tang*, 1989; *Zhang*, 1989; *Chen et al.*, 1991; *Cowie et al.*, 1993].

**2.2. Tests of SOC: Observable Properties of the Assumed SOC State of the Crust**

What are the possible observable consequences of the SOC hypothesis for the crust? A hypothesis cannot be tested by the empirical evidence that served to shape it. We thus exclude the Gutenberg-Richter power law as well as the fractal geometrical structure of the earthquake epicenters and fault patterns. In the present discussion, we address two novel properties/predictions that derive naturally from detailed numerical studies of simplified models of the crust.

The first one is the most obvious for geologists but we nevertheless address it as some confusion might exist, probably seeded by the statistical physics community. It concerns the localization of earthquake activity on faults [*Knopoff,* 1996]. When elasticity is correctly incorporated, SOC (defined by the four conditions above) is found to coexist (and is, in fact, deeply intertwinned) with a spontaneous organization of a fault structure on which the earthquake activity is clustered [*Cowie et al.*, 1993; *Miltenberger et al.*, 1993; *Sornette et al.*, 1994; *Sornette et al.*, 1995; *Sornette and Vanneste*, 1996]. SOC is thus not synonymous with a diffuse "avalanche" activity covering uniformly all the available space, as extrapolations from sandpile models would imply [*Bak and Tang*, 1989; *Chen et al.*, 1991; *Olami et al.*, 1992; *Christensen and Olami*, 1992]. The incorporation of elasticity in models of SOC, in fact, leads to an enrichment of the concept, since fault structures are found to be geometrical objects themselves submitted to the self-organizing principles [*Miltenberger et al.*, 1993; *Sornette et al.*, 1994] and which can be perceived as self-organized structures solving a global optimization problem as defined by *Crutchfield and Mitchell* [1995].

The most interesting aspect of SOC is probably in its prediction that the stress field exhibits long-range spatial correlations as well as important amplitude fluctuations. The exact solution of a simple SOC model [*Dhar and Ramaswamy*, 1989; *Dhar*, 1990; *Majumdar and Dhar*, 1991] has shown that the spatial correlation of the stress-stress fluctuations around the average stress is long range and decays as a power law. A similar conclusion on the related strain field fluctuations has been obtained within a general mathematical formalism [*Sornette and Virieux*, 1992]. The conclusion we can draw from the understanding brought by these conceptual models is that the stress fluctuations not only reflect but also constitute an active and essential component of the organizing principle leading to SOC. It is an intringuing possibility whether the observed increase of long-range intermediate-magnitude earthquake activity prior to a strong earthquake might be a witness of these long-range correlations [*Knopoff et al.*, 1996]. Figure 2 illustrates, in a numerical model discussed below, the existence of long-range spatial correlations between domains which are close to rupture. This correlation is represented geometrically by the large size of the connected domains in dark which are within 10% from rupture.

Two important consequences can be drawn. First, a substantial fraction of the crust is close to rupture instability. Together with the localization of seismicity on faults, this leads to the conclusion that a significant fraction of the crust is susceptible to rupture, while presently being quiescent. The quantitative determination of the susceptible fraction is dependent on the specificity of the model [*Zhang*, 1989; *Pietronero et al.*, 1991; *Sornette et al.*, 1994] and cannot thus be ascertained with precision for the crust. What is important, however, is that the susceptible part of the crust can be activated with relatively small perturbations or by modification of the overall driving conditions. This remark leads to a straighforward interpretation of induced seismicity by human activity as we discuss below.

Second, if a finite fraction of the crust is susceptible and can easily be brought to an unstable state, not all the crust is in such a marginal stability state. In fact, the complementary finite fraction of the crust is relatively stable and resistant to perturbations. The assertion often found in the SOC literature that "the crust is almost everywhere on the verge of rupture" is simply wrong, as found in the simplified SOC models. For instance, numerical simulations show that in discrete models made of interacting blocks carrying a continuous scalar stress variable, the average stress is around 0.6 times the threshold stress at rupture. In these models, the crust is far, on the average, from rupture. However, it exhibits strong fluctuations such that a finite subset of space is very close to rupture as already pointed out. The average is thus a poor representation of the large variability of the stress amplitudes in the crust. This leads us to predict that not all human perturbations will lead to induced seismicity and that some regions will be very stable against external disturbances.

Putting precise numbers and predicting the specific spatial domains that can be brought to instability is out of reach of present models which are not detailled enough to capture the rich variability of the Earth's crust. However, we can define workable quantities that can be measured and tested, such as spatial correlation functions, and we believe that the qualitative properties of the SOC state are already worthy of attention and can be compared with observations. We now turn to the available evidence for induced seismicity that can bring further light to our hypothesis.

## 3. Induced Earthquakes as Stress Gauges

### 3.1. Various Available Stress Gauges Testing the State of the Crust

In order to test our hypothesis on the long-range correlation of stress in the crust and its significance for induced seismicity, we need stress indicators. Let us first put this problem in perspective to appreciate better the innovation brought by induced earthquakes. Six such stress indicators are usually used to infer the orientation and the relative magnitude of the contemporary in situ tectonic stress field in the Earth's lithosphere: earthquake focal mechanisms, well bore breakouts, hydraulic fracturing and overcoring, and young geologic data including fault slips and volcanic alignments [e.g., *Zoback and Zoback*, 1980]. These stress gauges provide some indication on the orientation of the stress tensor and only on the relative magnitude of the stress components [e.g,. *Gephart and Forsyth*, 1984]. The hydraulic fracturing stress measurement is the only technique that provides information on horizontal stress magnitudes and orientations at the same time [e.g., *Haimson and Fairhurst*, 1970; *Zoback and Haimson*, 1983].

There is a long history of induced seismicity. For instance, it has been suggested that earthquakes may be associated with the filling of artificial reservoirs [*Carter*, 1945; *Gupta and Rastogi*, 1976; *Simpson*, 1986] and with fluid injection under pressure greater than hydrostatic [*Evans*, 1966; *Healy et al.*, 1968; *Raleigh et al.*, 1972]. Reservoir impoundments trigger seismicity either by direct loading effects or by coupled poroelastic effect [*Simpson*, 1986]. The direct effects of the load can be divided into an elastic effect and an undrained poroelastic effect. The elastic effect modifies the state of stress in accordance with the tectonic environment and changes the seismicity rate as can be deduced from the Coulomb failure envelope, for instance, inducing seismic activity in normal faulting areas and leading to seismic quiescence in thrust faulting areas [*Bufe*, 1976; *Jacob et al.*, 1979]. The second direct load effect, the increase of pore pressure immediately below the reservoir, always favors induced seismicity. In this way, many seismic events with magnitudes 3 to 4 have been reported in the vicinity of subsurface fluid injections, e.g. hydrocarbon fields stimulated by fluid injection or waste storage (Colorado [*Evans,* 1966; *Healy et al.*, 1968; *Raleigh et al.*, 1972], New York, [*Fletcher and Sykes,* 1977], Nebraska [*Evans and Steeples*, 1987], and Ohio [*Nicholson et al.*, 1988]). Delayed seismic responses to reservoir impoundment correspond to a more gradual diffusion of water from the reservoirs to hypocentral depths [*Zoback and Hickman*, 1982; *Simpson D.W. et al.*, 1988; *Ferreira et al.*, 1995]. Similar possible connections between fluid manipulations and earthquakes at distances of tens of kilometers were also proposed at Denver and Lacq (France) fields on the basis of pore pressure diffusion associated with deep well activities [*Hsieh and Bredehoeft*, 1981; *Grasso et al.*, 1992a]. Because of the lack of knowledge of hydrological properties at depth, the mechanisms leading to a delay of induced seismicity are not constrained enough to derive quantitative estimates of the induced perturbations. Accordingly, we do not use the events triggered by the pore pressure diffusion mechanism even if they constitute an important problem for future investigations. One must note that there is still a big gap between the classification of relevant reservoir-induced seismicity mechanisms and examples that more often appear as complex coupled processes than really succeed in isolating one of these mechanisms.

Earthquakes induced by fluid loading are understood to result from an increase in pore pressure, causing an effective normal stress reduction on the fault plane. This simple explanation leads to a paradox when applied to the numerous cases where hydrocarbon extractions triggered seismic activity [*Grasso*, 1993]. From the observations of seismic events recorded by local networks in areas surrounding hydrocarbon fields, e.g., in Alberta, Canada [*Wetmiller,* 1986], in SW France [*Grasso and Wittlinger*, 1990; *Guyoton et al.* 1992], and in northern Holland [*Haak,* 1991], it appears that such earthquakes have not occurred in the depleted rock reservoir where gas pressure has decreased but instead occurred above or below the gas reservoir. These observations are explained by a poroelastic stress transfer from the depleted reservoir to the levels surrounding the hydrocarbon reservoir [*Segall,* 1989, 1992]. The linear relationship between the decrease in pore pressure and the value of surface subsidence near the Lacq gas field [*Grasso et al.*, 1992b] supports the poroelastic modeling of the local strain and stress fields. At other sites where the available measurements do not permit the unambiguous identification of the rock mass rheology, *Grasso* [1992] proposed a two-step mechanism in which poroelastic-inelastic behavior takes place before the in situ faults are reactivated. In these cases, earthquakes ($M_{max}$~4) are a consequence of a principally aseismic displacement. In a broader context, massive hydrocarbon recovery reduces the vertical load which can trigger earthquakes in a thrust faulting environment. The same basic relation explains the energy that is released by earthquakes induced by mining [*McGarr*, 1976]. *McGarr* [1991] suggests a mechanical connection between the mass withdrawal and the earthquake size: the seismic deformation (estimated from the seismic moment) is that required to offset the force mismatch caused by hydrocarbon production which is proportional to mass withdrawal. The other types of seismicity induced by mass withdrawal include those due to quarrying [e.g., *Pomeroy et al.*, 1976; *Yerkes et al.*, 1983] and those associated with the removal of ice sheets [*Stein et al.*, 1979; *Johnston*, 1987, 1989; *Arvidson*, 1996].

### 3.2. Stress Variations From Pore Pressure Changes

As already mentioned, a reliable measure of absolute stress is very difficult. In addition, it is probably useless for our understanding the earthquake phenomenology. It is rather the distance between the local stress to the rupture threshold that informs us on the organization of the crust and indicates its susceptibility to rupture. The crust is a heterogeneous medium, characterized by varying properties in space and also in time. Suppose that we were to measure a small stress at some location. Should we conclude that this location is not susceptible to exhibit earthquakes? Certainly not, as this location might be relatively weak and the stress threshold for rupture might be only slightly larger. This paradox has been discussed, for example, in the case of the San Andreas fault [*Wang and Sun,* 1990; *Zoback and Healy*, 1992; *Saucier et al.*, 1992; *Zoback*, 1993; *Scholz and Saucier,* 1993]. Figures 1 and 2 show the map of absolute stress and of stress over local rupture threshold, respectively, in a simple antiplane elastic model of the crust sheared on the upper and lower boundary. This model has the SOC properties and shows a self-organization of faults (see *Cowie et al.* [1993], *Miltenberger et al.* [1993], and *Sornette et al.* [1994] for details). Comparing these two maps, we can see that locations where the stress is strong does not systematically imply rupture. Reciprocally, a region with a small stress may nevertheless be very close to rupture. Figure 3 shows the actual faults on which all the earthquake activity is localized. We see that a large domain where earthquakes never occur has its stress very close to its rupture threshold, implying that tiny perturbations might destabilize and induce seismicity in this domain.

A very interesting and important aspect of induced seismicity is that it provides a measure of the distance of the local stress to the rupture threshold if one is able to quantify the stress variation induced by human activity. Since this stress variation varies progressively in time, one can in principle detect the amount and nature of stress needed to trigger seismicity. This amount gives an estimation of the distance to rupture prior to the human activity. We now present a short review on the determination of stress variations that have caused induced seismicity.

An upper bound of the amplitude of the pore pressure variation necessary to cause reservoir-induced seismicity (RIS) is given by the water height of the artificial surface reservoir. The most common values are around 1-2 MPa for most of $M \geq 2.5$ RIS [see, e.g., *Gupta*, 1985, Table 1]. In some areas, fluctuations of a few meters of water level correlate with seismicity, in agreement with the model of a coupled pore pressure field beneath the reservoir [*Simpson and Negmatullaev*, 1981; *Roeloffs*, 1988]. *Roeloffs* [1988] argues that a 0.1 MPa stress step is needed to trigger seismicity. Similarly, upon fluid injection performed in salt recovery in Dale, New York, a monthly rate of 80 earthquakes was observed against a background of one natural earthquake per month [*Fletcher and Sykes,* 1977]. Downhole pressures ranged from 5.2 to 5.5 MPa at 450 m ($P_{litho}/P_{fluid} \approx 1.9$), but for pressures below 5 MPa, there was no seismic activity. One must note that when evidence for subhydrostatic preexisting pore pressure is reported at depth, the pore pressure change at depth can be larger than that at the surface. This is what happened at Denver. The pore pressure change at depth was over 10 MPa, but the head change at the surface was essentially zero (the waste simply flowed into the well). In the context of RIS, pore pressure at hypocentral depth may be below the hydrostatic value before impoundment; therefore pore pressure increase may be more than deduced from simple loading evaluations.

In the case of the Lacq gas field, hydromechanical parameters are constrained by numerous subsurface data such as porosity, bulk and elasticity moduli, Biot coefficient, pore pressure history, and surface subsidence. Using the value of these parameters estimated in situ, the application of the poroelastic stressing model to the Lacq gas field predicts that a variation in shear stress of a few bars triggers seismicity above and below the gas reservoir [*Segall et al.*, 1994]. In this case, we can make an order-of-magnitude estimate of the expected stress changes using [*Segall,* 1989, 1992; *Segall et al.*, 1994]

$$\Delta\sigma_{max} \approx [4\mu/(1-\nu)\pi D] \ \Delta H_{max} \qquad (1)$$

where $\mu$, $\nu$, $D$, $H_{max}$, are the shear modulus, the Poisson coefficient of the rock matrix, the depth of reservoir, and the maximum subsidence depth, respectively. At other sites, we estimated the maximum induced stress at the onset of seismicity either using the subsidence and shear modulus of the rocks involved or using an approximate determination based on pressure drop and reservoir geometry as proposed by *Grasso* [1992]. In this case, we get

$$\Delta\sigma_{max} \approx [(1-2\nu)/2\pi(1-\nu)] \ \Delta p \ F_{max}(a/D), \qquad (2)$$

or

$$\Delta\sigma_{max} \approx 0.1 \ (T/D) \ \Delta p \ F_{max}(a/D), \qquad (3)$$

for standard (limestone) reservoirs. $\Delta p$ and $F_{max}(a/D)$ are the reservoir pressure drop and a dimensionless function of position and reservoir geometry, respectively, $a$ is the reservoir radius, $T$ is the reservoir thickness, and $D$ is the reservoir depth (the maximun value is $F_{max}(1) \approx 3$, [*Segall*, 1989]). For all the sites where estimates can be performed, stress changes smaller than 1 MPa appear to trigger seismic instabilities (Table 1). Seismic instabilities occur under a great variety of regional tectonic conditions: stable in the Netherlands and Texas; stable and/or extension tectonics in the North Sea, Norway and Denmark; compression tectonics in the Caucasus, France, and Canada. In Table 1a, we summarize the parameters for the seismicity induced by these deep well pore pressure fluctuations. One must note that due to the poroelastic effect, the induced stress change is much smaller than the change in bottom hole pressure (BHP) relative to the initial fluid pressure (Table 1a). Thus the data presented by *Nicholson and Wesson* [1993] must be read accordingly. Both increases or decreases of pore pressure have triggered $M< 5$ earthquakes that are located a few kilometers away from well activities [*Grasso*, 1992; *Nicholson and Wesson*, 1993]. When secondary recovery is performed using fluid injection on the periphery of the depleted reservoirs, a coupling between the two mechanisms generates new fractures [e.g., *Davis and Pennington*, 1989]. In these cases, the estimates of the critical induced stress are not accurate, and we do not use these data.

### 3.3. Stress Variations From Hydrocarbon Extraction

The stress variations that trigger seismicity in the neighborhood of hydrocarbon extraction appear to be approximately of the order as those estimated to trigger seismicity in the vicinity of artificial water reservoir impoundments. Some major earthquakes ($M \geq 6$) may be related to hydrocarbon production. Both in California [*McGarr*, 1991] and in Uzbekistan [*Grasso*, 1993], major events can be related to upper crustal hydromechanical disturbances produced by subsurface fluid extraction from shallow geological formations. Despite the smaller fluid pressure changes observed at these sites compared to those previously described to be the cause of moderate ($M<5$) earthquakes by pore pressure effect, larger seismic responses ($M \geq 6$) located at greater distance from the hydrocarbon reservoir are observed [*McGarr*, 1991; *Grasso,* 1993]. The triggered earthquakes occurred in deep fault zones buried under anticlines containing hydrocarbon reservoirs. Following the analysis of *McGarr* [1991] that mass withdrawal at these sites is related to the major earthquake sequences, we can obtain an estimate of stress changes for each sequence. We multiply the seismic stress drop by the ratio of the loading system stiffness (the vertical force change) to the stiffness of a circular crack model. The results are presented in Table 1b. Although less direct and reliable than the values extracted from the pore pressure effects, these estimates suggest that earthquakes can be triggered by small shear stress perturbations (<1 MPa). Note that for the M≥6 earthquakes that occurred in the Gasli field area, the orientations of the *P* axes of maximum compression span nearly half (i.e., 90°) of their maximal range (180°) (Figure 4). Such abnormal variations in orientation can either indicate a local stress anomaly induced by the hydrocarbon production or provide evidence for fluid contributions to fracturing, i.e. weak faults stimulated by tight fluid seals that decrease the apparent friction coefficient [*Blanpied et al.,* 1992].

A typical fluid removal onthe order of magnitude of $10^{11}$ kg (Table 1b) appears to be sufficient to induce large thrust events in a compressive setting. This leads to two remarks. First, the same mass threshold has been reached in other hydrocarbon fields where $\sigma_1 \neq \sigma_H$ and for which the poroelastic response by itself has induced only moderate seismicity (*M*<5) close to the depleted reservoirs (e.g. Lacq field western Pyrénées foreland; Groningen field northern Holland; Ekofisk field, central graben Norway, Table 1a). In agreement with the admitted tectonic stress field for the three areas given as examples [e.g., Zoback 1992], the lack of thrust faulting in these three regions has not permitted the generation of large seismic failures in response to mass withdrawal. In the absence of thrust faulting, only the poroelastic stressing effect, which is independent of the preexisting tectonic stress, can induce earthquakes. On this basis, we propose that the comparison between seismicity triggered by poroelastic stress change and by mass transfer may be used to infer the local tectonic state of stress. When the single poroelastic response is observed, the location and geometry of volumes brought toward failure by the stress change are coupled to the applied boundary conditions, i.e., to the orientation of the regional tectonic stresses [*Segall et al.*, 1994]. In this way, we can recover the tectonic state when the stress field anisotropy is small. Larger anisotropy of the stress field, i.e., when the horizontal stress is larger than the vertical stress, will allow isostatic rebound for mass withdrawal and conversely.

Second, the estimated value of the critical mass withdrawal necessary to induce thrust events is at the lower limit of the range of values of the increases in load that trigger fast normal faulting response in reservoir impoundment [*Gupta and Rastogi*, 1976; *Simpson D.W. et al.*, 1988]. Accordingly, normal faulting in reservoir-induced seismiciy also is an indicator that confirms the regional extension setting of the African midplate stress field (Kariba), the local extensional state of deformation of different foot hills massifs (Sierra Nevada, Oroville; French subalpine massif, Monteynard), and the western Anatolian province (Kremasta). These earthquakes are the only ones with normal faulting source mechanism among the top nine with M>5.0 RIS listed in Table 2. Note that in thrust regions, load increases inhibit the occurrence of thrust earthquakes. Accordingly, numerous reservoir impoundments, for water height ≥ 100 m, have not triggered fast seismic response within Himalayan thrust areas [*Gupta*, 1985]. For the other fast seismic responses to reservoir impoundments, the observed reverse-slip or strike-slip events are associated with local increase of pore pressure in undrained conditions that are independent of the tectonic setting [*Gupta and Rastogi,* 1976; *Simpson*, 1976, 1986; *Gupta,* 1985]. Whatever the mechanism of pore pressure change (drained or undrained), the water height is the key parameter that determines the stress change necessary to trigger seismicity.

Despite the accepted relatively uniform minimum value of 100 m for the water height thought to induce seismicity [e.g,. *Rothé*, 1970], there is no apparent correlation between water height and the size of largest RIS (Figure 5). The Brazil RIS example [*Ferreira et al.*, 1995] is better than most cases in showing a clear relationship between seismicity and water level, but the statement is weakened by the observation of a delayed correlation between seismicity and RIS already for a water depth of only 30 m (Figure 6). These observations support the hypothesis that the sizes of the reservoir-induced-earthquakes are more controlled by the organization of the crust at large scales than by the size of the perturbation produced by the loading system. This is clearly seen in the center of Figure 2 for the model of earthquakes already mentioned above: no earthquake ever occurs in this central region. However, if one was to increase slightly the local stressby using an external source so as to reach the rupture threshold, the type of seismicity in this model would be of the same nature as on the self-organized faults shown in Figure 3 in the absence of the external source. The reason is that the correlation in the stress field seen in figure 2 is of the same statistical nature both close to the faults and far from them. Direct numerical simulations confirm this hypothesis. We thus view the evidence in Figure 2 as compatible with and even strengthening the SOC hypothesis. It is not easy to imagine another framework that could rationalize all these different observations.

### 3.4. Properties of Induced Seismicity

After the inception of induced seismicity, it is often observed that much weaker variations in stresses (~0.1 MPa), related to variations either in surface water level storage (Figure 6) or in subsurface fluid pressure drop (Figure 7), control for many following years the ongoing seismic slips occurring on preexisting discontinuities. In several areas, smooth and slow stress/strain changes, at least 1 order of magnitude smaller than the triggering stress variation, sustain the induced seismicity over months or years. This should be compared to the small stress transfer (~0.1 MPa) in naturally activated seismicity that has been proposed to explain coseismic slip-triggered or slip-locked when loaded by an earthquake on neighbouring faults [*Jones et al.*, 1982; *Simpson R.W. et al.*, 1988] in areas prone to failure. Recent models of coseismic effects of the Landers, California, earthquake support that a 0.1-1 MPa stress change affects seismicity 30-100 km away from the main $M$~7.4 shock [*Harris and Simpson*, 1992; *Stein et al.*, 1992; *Jaumé and Sykes*, 1992]. These effects may be due to dynamic strains associated with seismic waves or with magma chambers [*Hill et al.*, 1993]. We can also draw an analogy with the seismic swarms that are activated repetitively on volcanoes by moon tides [*Rydelek et al.*, 1988]. In these cases, the minimum stress changes that induce seismicity have been estimated to be a few hundredths of a megapascal. Although the search for RIS/tide correlation is difficult due to the limited available data, tidal stress triggering of RIS has been proposed as significant in a few sites [*Klein*, 1976]. For the cases reviewed in the present study, changes in stress of a few hundredths of a megapascal. are sufficient to modify previous risk assessments, based on natural seismicity (e.g., northern Holland; Aquitaine basin, France; SW India, SE China; western Uzbekistan [*Grasso,* 1992]. The very small value of the stress variations necessary to sustain seismicity is in agreement with the critical state hypothesis (see Figures 1-3). These observations give us a theoretical basis for the probable existence of correlations between seismicity and tides in certain circumstances.

Volcanic activity represents another class of induced seismicity which has been studied recently in the context of SOC [*Diodati et al.*, 1991; *Sornette,* 1992; *Grasso and Bachelery*, 1995]. Let us mention that the endemic seismicity of the Piton Fournaise volcano exhibits both a power law distribution of event sizes and a strong diurnal behavior that underlines the sensitivity of the system to weak external forcing (Figure 8).

In mining induced seismicity, the seismic events disappear during weekends and holidays (Figure 9). Within a few days, the seismicity stops in contrast to what happens for reservoir and gas extraction induced seismicity. The difference can be explained by the fact that the stress variation necessary to trigger seismic events is relatively large (a few tens of megapascals), by comparison to the reservoir and gas extraction cases. Within our SOC hypothesis summarized in Figures 1-3, we are led to interpret this case as corresponding to a location where the stress is relatively far from the rupture threshold, as in the light grey areas in Figure 2. Therefore a large external stress must be applied to reach the rupture threshold. Suppose that an earthquake is then triggered and relaxes the stress by elastic redistribution. The point is that in contrast to areas close to rupture, most of the area around the induced earthquake is itself in a state of low stress relative to its threshold. This neighboring region will thus act as a sink of stress and no other earthquake will be triggered if the external stress source is deactivated. Note that this argument relies on the long-range correlation of the stress field, already discussed above. The rate of mining extraction corresponding to a front progress of a few meters per day gives an upper bound for the external stress rate applied to the system. We exclude an alternative explanation that the difference is that failure is reached in mining by increasing the shear stress and by reducing the effective normal stress in RIS. The oil and gas induced seismicity examples are reported to be located above and below the reservoir levels where pressure and mass withdrawals occur [*Grasso*, 1993]. Typical poroelastic stress changes induce stress transfers outside reservoir levels [*Segall et al.*, 1994]. There is no fluid circulation between the reservoir level and the upper level. Accordingly, the earthquakes are not triggered by reducing the effective normal stress as popularized by the Denver injection case study.

It is important to stress that not all mining, reservoir impoundment, and hydrocarbon recovery lead to induced seismicity [*Grasso*, 1993]. There exist strong stress perturbations that have not led to any seismicity. For the same distance range, regional effects of fluid manipulation are reported either to increase or to decrease the regional seismicity depending on the sign of the fluid pressure variations [*Grasso et al.*, 1992a]. In the Himalayan thrust region, load increase inhibits the occurrence of thrust earthquakes. Similarly, numerous reservoir impoundments, for water height ≥ 100 m, have not triggered seismic response [*Gupta*, 1985]. This is not in contradiction to the proposed physical picture. Indeed, in order to induce seismicity, a perturbation must be of the correct sign and in the correct tensorial direction. It is clear that the elastic effect induced by fluid loading will change the radius of the Mohr circle depending of the tectonic environment and will modify the seismicity accordingly with respect to the Coulomb failure envelope: this leads to an induced seismic activity in normal faulting area and to induced seismic quiescence in thrust faulting areas [*Bufe*, 1976; *Jacob et al.*, 1979]. In the simplified models of SOC, this last situation would be represented by an effective local deloading of the system. The example of the Himalayan thrust region illustrates the importance of the direction of the stress tensor perturbation in comparison with that of the natural stress field. Another vivid example is provided by the South American subduction belt for which no induced seismicity has been observed as a response to the fluid impoundments in a large number of reservoirs. Here again, a load increase inhibits thrust earthquakes to occur and stabilize the system. As already pointed out, these observations are in agreement with the large variability of the stress field and rupture conditions that characterize the SOC state.

We note finally that in all cases where a monitoring has been done [e.g., *Grasso*, 1992], the induced seismicity is characterized by a power law frequency-size distribution (Figure 10), similar to the Gutenberg-Richter distribution. This is also expected from our SOC hypothesis as explained above in relation to Figures 1-3.

9## 4. Maximum Earthquake Magnitude and Finite Size Effects

The largest induced events ($m_0 \geq 10^{18}$ N m) triggered either by water reservoir impoundements or hydrocarbon recoveries are located in area where thick seismogenic layers are found close to the surface, i.e., in shield areas. This is observed for the top four induced events that occurred in low-seismicity areas: Gazli, central Asia; Hsinfengian, southeast China; Kariba, Zambia-Zimbabwe; Koyna, southwest India. This would suggest that shield earthquakes could be breaking the entire brittle crust. Because of the slow rate of seismic energy released in a shield area, this hypothesis is difficult to check. In the Gazli area, we have observational evidence that the earthquakes occur at depths from 0 to 25 km [*Pearce*, 1987; *Bowers and Pearce*, 1995; *Bossu et al.*, 1996]. In other areas of reported induced seismicity with *M*>3, the thickness of the larger brittle sedimentary bed (for seismicity induced by impoundment or depletion of reservoir) or the geomechanical setting (dimension of the mine pillars in the case of mining) seems to provide a rough upper scale that limits the size of the largest possible earthquake. These characteristic scales are significantly smaller than for shield areas, and the largest earthquakes are smaller, as expected. We should, however, mitigate these ideas as there are also examples, for instance, in Brazil, where RIS in cratonic areas contains only microearthquakes. This is due to the fact that the rheology (controlled by the temperature, the material strength, and the rate of deformation) is also an important ingredient, in addition to the characteristic scales, including the time scale for observation. Another possibility is that the induced seismicity could be subcritical. In this case, seismicity will not be limited so much by geometrical constraints such as a finite seismogenic depth [*Kagan*, 1997; *Main*, 1996].

Persuing this idea further, *Scholz* [1982] proposed that the size of the largest earthquakes should be controlled by a characteristic dimension of the whole brittle crust. Several authors have used the size of the rolloff in the Gutenberg-Richter distribution as a measure of the characteristic scale of the brittle crust [e.g., *McGarr*, 1986; *Shimazaki*, 1986; *Pacheco et al.*, 1992; *Romanowicz*, 1992; *Romanowicz and Rundle*, 1993]. However, *Sornette et al.* [1996] have shown that the quantitative use of this correspondence is doubtful. Indeed, the accuracy of the correlation between the rolloff size of the frequency-size distribution and the width of the seismogenic zone is much worse than usually assumed and may be spoiled by interfering mechanisms [*Sornette and Sornette*, 1994]. It can at best give an order of magnitude of the thickness. In any case, we note that the characteristic scale, if any, should be easier to determine by induced earthquakes than by natural earthquakes for which the relationship between structure and processes is complex and the rheology is poorly constrained.

In this spirit, the upper size of the earthquakes is roughly in agreement with the break in self-similarity (see Figure 11 for examples on recurrent break in slope of self-similarity). The correlation can be improved when using several lines of evidence put together, such as a rolloff of the frequency-size distribution and a rolloff of the log-log plot measuring the fractal dimension the set of earthquake epicenters [*Volant and Grasso*, 1994]; (see *Ouillon et al.* [1996] for an implementation of this dual strategy using multifractal and wavelet analysis for tectonic faulting). When attempting to correlate the rolloffs in the earthquake frequency-size distribution with the characteristic scale of the geomechanical setting (Figures 12-14), it becomes clear that the relationship is much more complex than assumed initially as for instance by *McGarr* [1986], *Shimazaki* [1986], *Pacheco et al.* [1992], *Romanowicz* [1992], Romanowicz and Rundle [1992]. This is true for the characteristic size (*M*~3) observed for seismicity induced by ore mining, by depleting a hydrocarbon reservoir, by impounding an artificial lake and by building a dyke on a basaltic volcano as reported in Figure 12, 13, 14 and 8a, respectively [*Grasso*, 1993; *Grasso and Bachelery*, 1995]. Both geometrical and dynamical characteristics of the driving processes allow a large range of characteristic sizes. For volcanic earthquakes for instance, *Grasso et al.* [1994] show that $M_c$ span from less than *M*~2 at Merapi volcano to *M* ~5 at Mount St. Helens. This is to be compared with the close similarity of the thicknesses of their respective seismogenic layers which are bounded by the shallow magma reservoir depths corresponding to an order of magnitude of a few kilometers. Depending on the eruption process, either the whole thickness of the seismogenic bed is activated (explosion type [*Main*, 1987] and accordingly defines the largest critical size for the induced earthquakes, or the size and viscosity of the magma intrusion defines a smaller critical size for the earthquake distribution [*Grasso et al.*, 1994].

## 5. Concluding Remarks

Major reported cases of induced seismicity in various parts of the world have been reviewed. These studies provide direct evidence that both pore pressure changes ($\pm\Delta p$) and mass transfers ($\pm\Delta m$), corresponding to incremental differential stresses of the order of 1 MPa regardless of the tectonic setting, trigger seismic instabilities in the uppermost crust with magnitudes ranging from 3.0 to 7.0. Moreover, with stress steps with amplitude around 0.1 MPa, at least 1 order of magnitude smaller than the trigger step, sustained induced seismic activity is often observed over long period of times (Figures 6 and 7).

We have proposed to view the Earth's crust as functioning in a self-organized critical state. The evidence provided by induced seismicity has been argued to be consistent with the SOC hypothesis. For instance, a prediction from SOC is that a finite fraction of the crust may be awakened in response to a perturbation, as a result of the subtle spatial correlations of the stress field which characterize the self-organized critical state. We have attempted to clarify the notion of self-organized criticality and to solve some apparent contradictions between this concept and some observations. In particular, the importance of the sign and tensorial nature of the perturbation, in comparison with the local tectonic stress field, has been emphasized. This explains in a natural way why not any perturbation can induce seismicity.

We have also suggested that the thickness of the largest local seismogenic bed provides a rough indication of the characteristic scale up to which self-similarity can be observed. This characteristic dimension mimics the effect that the schizosphere plays for tectonic earthquakes [*Scholz*, 1991. In induced seismicity, it ranges from the pillar size in mining to the thickness of brittle sedimentary beds in the vicinity of reservoir and gas extraction setups. However, we have cautioned and discussed examples where this correspondence does not work well due to the existence of other physical mechanisms than just the finite size effect which can compete and modify the dynamics. We propose that sedimentary basins with soft material, which are prone to local destructive amplification of seismic waves, are less prone to strong shallow earthquakes. Conversely, thick cratonic or shield areas have localized the largest induced earthquakes ($M \geq 6$) and are thus candidates for future large shallow induced earthquakes. The identification and study of such "isolated geological objects" with usually low seismicity rates (e.g., as in intraplate areas) can help in the estimatation of the size of the largest possible earthquake. In terms of seismic risks, estimates of earthquake sizes, which use the characteristic dimension of local induced earthquakes, can provide a complementary view point to the analysis of *Boatwright and Choy* [1992] who focused on acceleration source spectra of large cratonic earthquakes. The properties of induced seismicity and their rationalization in terms of the SOC concept suggest that potential seismic hazards extend over a much larger area than that where earthquakes are frequent, as proposed by *Kagan* [1997].

An important idea defended and documented in this paper is the possible existence of large-scale spatial correlation of the stress field in the crust. This might have important use for the detection and use of earthquake precursory phenomena [*Dobrovolsky et al.*, 1979; *Sornette and Sammis*, 1995; *Knopoff et al.*, 1996; *Saleur et al.,* 1996] if any (see *Main* [1996] for a review). We note also that small stress changes, with the same amplitude range as the ones that sustain induced seismicity, can explain large distance anomalies before some earthquakes associated with hydrological changes [e.g., *Roeloffs*, 1989; *Silver and Valette-Silver*, 1992] and electrical changes [ e.g., *Bernard*, 1992] and are now proposed to be driven by local fluid-induced instabilities.

**Acknowledgements:** One of us (J.-R. G.) thanks D.W. Simpson for discussions on RIS as well as A. McGarr and P. Volant for fruitful comments. P. Molnar, J.L. Chatelain, and O. Coutant provided helpful comments on an earlier version of the manuscript. This study benefited from the open collaboration with petroleum companies. Cag, Elf, Nam, Petroland, Phillips, and Shell provided geomechanical data of the fields studied in this work, but the interpretations expressed in this study are the authors' own. This research was partially supported by GdR mécanique des roches profondes, the Elf company under contract Risk reassesment within the Lacq Industrial Facilities and the French DBT-INSU Instability Program. This work is publication No 4883 of the Institute of Geophysics and Planetary Physics, University of California, Los Angeles.

# References

Amorese, D., and J.-R. Grasso, Rupture planes of the Gazli earthquakes deduced from local stress tensor calculation and geodetic data inversion: Geotectonic implications, *J. Geophys. Res.*, *101*, 11,263-11,274, 1996

Amorese, D., J.-R. Grasso, L.M. Plotnikova, B.S. Nurtaev, and R. Bossu, Rupture kinematics of the three Gazli major earthquakes from vertical and horizontal displacements data, *Bull. Seismol. Soc. Am.*, *85*, 552-559, 1995.

Arvidson, R., Fennoscandian earthquakes: Whole crustal rupturing related to postglacial rebound, *Science*, *274*, 744-746, 1996.

Bak, P., and C. Tang, Earthquakes as a self-organized critical phenomenon, *J. Geophys. Res.*, *94*, 15,635-15,637, 1989.

Bak, P., C.Tang and K.Weisenfeld, Self-organized criticality: An explanation of 1/f noise, *Phys. Rev A* , 38, 364-374, 1987.

Bernard, P., Plausibility of long distance electrotelluric precursors to earthquake, *J. Geophys. Res.*, *97*, 17531-17546, 1992.

Besrodny, E. M., The source mechanism of the Gazli earthquakes of 1976-1984, in Gazli Earthquakes of 1976 and 1984, Fan, Tashkent, pp.94-105, 1986.

Blanpied, M. L., D. A. Lockner, and J. D. Byerlee, An earthquake mechanism based on rapid sealing of faults, *Nature*, *358*, 574-576, 1992.

Boatwright, J., and G. L. Choy, Acceleration spectra anticipated for large earthquakes in NE North America, *Bull. Seismol. Soc. Am*., *82*, 660-682, 1992.

Bossu , R., and J.-R. Grasso, Stress analysis in the intraplate area of Gazli, Uzbekistan, from different sets of earthquakes focal mechanisms, *J. Geophys. Res.*, *101*, 17,645-17,659, 1996.

Bossu, R., J.-R. Grasso, L. M. Plotnikova, B. Nurteav, J. Frechet, and M. Moisy, Complexity of intracontinental seismic faultings: The Gazli, Uzbekistan, sequence, *Bull. Seismol. Soc. Am.*, *86*, 959-971, 1996.

Bowers, D., and R. G. Pearce, Double couple moment tensors for the 1976 Gazli aftershock earthquake sequence, *Tectonophyics*, *248*, 193-206, 1995.

Bufe ,C. G., The Anderson Reservoir seismic gap-Induced aseismicity?, *Eng. Geol.*, *10*, 255-262, 1976.

Burridge, R., and L. Knopoff, Model and theoretical seismicity, *Bull. Seismo. Soc. Am.*, *57*, 341-371., 1967.
Byerlee, J. D., Frictional characteristics of granite under high confining pressure, *J. Geophys. Res.*, *72*, 3639-3648, 1967.
Byerlee, J. D. and D. Lockner, Acoustic emission during fluid injection into rock, in *Proceedings of the first Conference on Acoustic Emission/Microseismic Activity in Geologic Structures and Materials, Penn. Stae Univ., June 1975,* edited by H. Hardy and F.W. Leighton, Trans Tech pp 87-98, Clausthal-Zellerfeld, Germany, 1977.
Carter, D. S., Seismic investigation in the Boulder dam area, 1940-1945, and the influence of reservoir loading on earthquake activity, *Bull. Seismol. Soc. Am.*, *35*, 175-192, 1945.
Carlson, J. M., J. S. Langer, B. Shaw, and C. Tang, Intrinsic properties of the Burridge-Knopoff model of a fault, *Phys. Rev. A*, *44*, 884-897, 1991.
Chen, K., P. Bak, and S. P. Okubov, Self-organized criticality in crack-propagation model of earthquakes, *Phys. Rev. A, 43*, 625-630, 1991.
Christensen, K., and Z. Olami, Variation of the Gutenberg-Richter *b* values and nontrivial temporal correlations in a spring-block model for earthquakes, *J. Geophys. Res*, *97*, 8729-8735, 1992.
Cook, N. G. W., Seismicity associated with mining, *Eng. Geol*., *10*, 99-122, 1976.
Costain, J. K., G. A. Bollinger, and J. A. Speer, Hydroseismicity: A hypothesis for the role of water in the generation of intraplate seismicity, *Seismol. Res. Lett.*, *58*, 41-63, 1987.
Cowie P., C. Vanneste, and D. Sornette, Statistical physics model of complex fault pattern organization and earthquake dynamics, *J. Geophys. Res.*, *98*, 21,809-21,821, 1993.
Crutchfield, J. P., and M. Mitchell, The evolution of emergent computation, *Proc. Na. Acad. Sci. U.S.A.*, *92*, 10,742-10,746, 1995.
Davis, S. D., and W. D. Pennington, Induced seismic deformation in the Cogdell oil field of west Texas, *Bull. Seismol. Soc. Am*., *79*, 1477-1494, 1989.
Deliac E. P., and N. C. Gay, The influence of stabilizing pillars on seismicity and rockbursts at ERPM, in *Proceeding of the 1st International Congress on Rockburst and Seismicity in Mines, Johannesburg, 1982,* edited by N.C. Gay and E.H. Wainwright , pp. 00-00, S. AFr. Inst. of Min. and Metall., Johannesburg, 1984.
Dhar, D., Self-organized critical state of sandpile automaton models, *Phys. Rev. Lett.*, *64*, 1613-1616, 1990.
Dhar, D., and R. Ramaswamy , Exactly solved model of self-organized critical phenomena, *Phys. Rev. Lett.*, *63*, 1659-1662, 1989.
Diodati, P., F. Marchesoni, and S. Piazza, Acoustic emission from volcanic rocks-An example of self-organized criticality, *Phys. Rev. Lett.*, *67*, 2239-2243, 1991.
Dobrovolsky, I. P., S. I. Zubkov, and V. I. Miachkin, Estimation of the size of the earthquake preparation zones, *Pure Appl. Geophys.*,*117*, 1025, 1979.
Evans, D. G., and D. W. Steeples, Microearthquakes near the Sleepy Hollow oil field, southwestern Nebraska, *Bull. Seismol. Soc. Am.*, *77*, 132-140, 1987.
Evans, D. M., Man-made earthquakes in Denver, *Geotimes*, *10*, 11-18, 1966.
Eyidogan, H., J. Nabelek and N. Toksoz, The Gazli, USSR, 19 March 1984 earthquake: The mechanism and tectonic implications, *Bull. Seismol. Soc. Am.*, *75*, 661-675, 1985.
Fabre, D., J.-R. Grasso, and Y. Orengo., Mechanical behaviour of deep rock core samples from a seismically active gas field, *Pure Appl. Geophys.*, *137*, 200-220, 1992.
Feignier, B., and J.-R. Grasso, Characteristics of seismic fractures as a function of geomechanical setting, *Pure Appl. Geophys.*, *137*, 175-199, 1992.
Ferreira, J. M., R. De Oliveira, M. Assumpsao, J. A. Moreira, R. G. Pearce, and M. K.Takeya, Correlation of seismicity and water level in the Acu reservoir: An example from NE Brazil, *Bull. Seismol. Soc. Am.*, *85*, 1483-1489, 1995.
Fletcher, J. B., A comparison between the tectonic stress measures in situ and stress parameters from induced seismicity at Monticelleo Reservoir, South Carolina, *J. Geophys. Res.*, *87*, 6931-6944, 1982.
Fletcher, J. B., and L. R. Sykes, Earthquakes related to hydraulic mining and natural seismic activity in western New York State, *J. Geophys. Res.*, *82*, 3767-3780, 1977.
Gephart, J. W., and D. W. Forsyth, An improved method for determining the regional stress tensor using earthquake focal mechanism data: Application to the San Fernando earthquakes sequence, *J. Geophys. Res.*, *89*, 9305-9320, 1984.
Gough, D. I., and W. I. Gough, Stress and deflection in the lithosphere near lake Kariba, *Geophys. J. R. Astron. Soc.*, *21*, 65-101, 1970.
Grasso, J.-R., Mechanics of seismic instabilities induced by the recovery of hydrocarbons, *Pure Appl. Geophys*., 139, 507-534, 1992.
Grasso, J.-R., Fluids and seismic instabilities: implication for the mechanical behavior of the uppercrust, doctorat d'état thesis, Univ. of Grenoble, Grenoble, France, 1993
Grasso, J.-R., and P. Bachelery, Hierarchical organization as a diagnostic approach to volcano mechanics: Validation on Piton de la Fournaise, *Geophys. Res. Lett.*, *22,* 2897-2900, 1995.
Grasso, J.-R., and G. Wittlinger, Ten years of seismic monitoring over a gas field area, *Bull. Seismol. Soc. Am.*, *80*, 450-473, 1990.
Grasso, J.-R., J.-P. Gratier, J.-F. Gamond, and J.-C. Paumier, Stress diffusion triggering of earthquakes in the upper crust, *Journ. Struct. Geol.*, *14*, 915-924, 1992a.
Grasso, J.-R., D. Fourmaintraux, and V. Maury, Le role des fluides dans les instabilités de la croute supérieure: L'exemple des exploitations d'hydrocarbures, *Bull. Soc. Geol. Fr.*, 163, 27-36, 1992b.
Grasso, J.-R., F. Guyoton, J. Fréchet, and J.-F. Gamond, Triggered earthquakes as stress gauges: Implication for risk re-assessment in the Grenoble area, France, *Pure Appl. Geophys.*, *139*, 579-605, 1993.
Grasso, J.-R., P. Bachelery, and A. Ratdomopurbo, Scaling of volcano eruptive styles by using induced self-organized seismicity, *Ann. Geophys.*, *12*, pp. C486, 1994.
Gupta, H. K., The present status of reservoir induced seismicity: Investigations with a special emphasis on Koyna earthquakes, *Tectonophysics, 118*, 257-279, 1985.
Gupta, H. K., and B. K. Rastogi, *Dams and Earthquakes*, 229 pp., Elsevier, New York, 1976.
Guyoton ,F., J.-R. Grasso, and P. Volant, Interrelation between induced seismic instabilities and complex geological structure, *Geophys. Res. Lett.*, *19*, 705-708, 1992.
Haak, H. W., Seismiche Analyse van de Aardbeving bij Emmen op 15 februari 1991, report, 14 pp., K. Ned. Meteorol. Inst., Minist. van Verkeer en Waterstaat, 14 pp., De Bilt, Nederlands, 1991.
Haimson, B. C., and C. Fairhurst, In situ stress determination at great depth by means of hydraulic fracturing, *Proc. U.S. Rock Mech. Symp.*, *11th*, 559-584, 1970.

Harris, R. A., and R. W. Simpson, Changes in static stress on southern California faults after the 1992 Landers earthquake, *Nature*, *360*, 251-254, 1992.

Healy, J. H., W.W. Rubey, D.T. Griggs, and C.B. Raleigh, The Denver earthquakes, *Science*, *161*, 1301-1310, 1968.

Hill, H. D., et al., Seismicity remotely triggered by the magnitude 7.3 Landers, California, earthquake, *Science*, *260*, 1617-1623, 1993.

Hsieh, P. A., and J.D. Bredehoeft, A reservoir analysis of the Denver earthquakes: A case of induced seismicity, *J. Geophys. Res.*, *86*, 903-920, 1981.

Ito, K., and M. Matsuzaki, Earthquakes as self-organized critical phenomena, *J. Geophys. Res.*, *95*, 6853-6860, 1990.

Jacob, K.H., J. Armbuster, L. Seeber, and W. Pennington, Tarbela reservoir, Pakistan: A region of compressional tectonics with reduced seismicity upon initial reservoir filling, *Bull. Seismol. Soc. Am.*, *69*, 1175-1182, 1979.

Jaumé, S.C., and L. R. Sykes., Changes in state of stress on the southern San Andreas fault resulting from the California earthquake sequence of April to June 1992, *Science*, *258*, 1325-1328, 1992.

Johnston, A.C., Suppression of earthquakes by large continental ice sheets, *Nature*, *330*, 467-469, 1987.

Johnston, A.C., The seismicity of "stable continental interiors," in *Earthquakes at North Atlantic Passive Margins: Neotectonics and Post Glacial Rebound*, edited by S. Gregersen and P.W. Basham, pp. 00-00, Kluwer, Norwell, Mass., 1989.

Jones, L.M., B. Wang, S. Xu, and T. J. Fitch, The foreshock sequence of the February 4, 1975, Haicheng earthquake (*M*=7.3), *J. Geophys. Res.*, *87*, 4575-4584, 1982.

Kagan, Y.Y., Seismic moment-frequency relation for shallow earthquakes: Regional comparison, *J. Geophys. Res.*, *102*, 2835-2852, 1997.

Klein, F.D., Tidal triggering of reservoir-associated earthquakes, *Eng.Geol.*,*10*, 197-210, 1976.

Knopoff, L., The organization of seismicity on fault networks, *Proc. Nat. Acad. Sci. U.S.A.*, *93*, 3830-3837, 1996.

Knopoff, L., T. Levshina, V.I. Keilis-Borok, and C. Mattoni, Increased long-range intermediate-magnitude earthquake activity prior to strong earthquakes in California, *J. Geophys. Res.*, *101*, 5779-5796, 1996.

Krantz, R. L., T. Satoh, O. Nishizawa, K. Kusunose, M. Takahashi, K. Masuda and A. Hirata, Laboratory study of fluid pressure diffusion in rock using acoustic emission., *J. Geophys. Res.*, *95*, 21,593-21,607, 1990.

Kusznir, N. J. and N. H. Al-Saigh, Some observations on the influence of pillars in mining-induced-seismicty, in *Proceeding of the 1st International Congress on Rockburst and Seismicity in Mines, Johannesburg, 1982,* edited by N.C. Gay and E.H. Wainwright , pp. 00-00, S. Afr. Inst. of Min. and Metall., Johannesburg, 1984.

Linde, A.T., et al., A slow earthquake sequence on the San Andreas fault, *Nature*, *383*, 65-68, 1996.

Lukk A.A. and S.L. Yunga, *Geodynamics and Stress Strain State of the Lithosphere of the Central Asia*, Donish, Duschanbe, Tajikistan, 1988.

Main, I. G., The characteristic earthquake model of the seismicity preceding the eruption of Mount St. Helens on 18 May 1980, *Phys. Earth Planet. Inter.*, *49*, 283-293, 1987.

Main, I. G., Statistical physics, seismogenesis, and seismic hazard, *Rev. Geophys.*, *34*, 433-462, 1996.

Majumdar, S.N., and D. Dhar, Height correlations in the Abelian sandpile model, *J. Phys. A Math. Gén.*, *24*, L357-L362, 1991.

McGarr, A., Seismic moments and volume changes, *J. Geophys. Res.*, *81*, 1487-1494, 1976.

McGarr, A., Some observations indicating complications in the nature of earthquake scaling, in *Earthquake Source Mechanics, Geophys. Monogr. Ser.*, vol. *37*, edited by S. Das, J. Boatwright, and C. H. Scholz, pp. 217-225, A.G.U., Washington, D.C., 1986.

McGarr, A., On a possible connection between three major earthquakes in California and oil production, *Bull. Seismol. Soc. Am*, *81*, 948-970, 1991.

McGarr, A., S. M. Spottiswoode, N. C. Gay, and W. D. Ortlepp, Observations relevant to seismic driving stress, stress drop and efficiency, *J. Geophys. Res.*, *84*, 2251-2261, 1979.

McGarr, A., R. W. E. Green, and S. M. Spottiswoode, Strong ground motion of mine tremors:Some implications for near source ground motion parameters, *Bull. Seismol. Soc. Am.*, *71*, 295-309, 1981.

Meade, R. B., Reservoirs and earthquakes, *Eng. Geol.*, *30*, 245-262, 1991.

Miltenberger, P., D. Sornette, and C. Vanneste, Fault self-organization as optimal random paths selected by critical spatio-temporal dynamics of earthquakes, *Phys. Rev. Lett.*, *71*, 3604-3607, 1993.

Nicholson, C., and R.L. Wesson, Triggered earthquakes and deep well activities, *Pure Appl. Geophys.*, *139*, 561-578, 1993.

Nicholson, C., E. Roeloffs, and R.L. Wesson, The northeastern Ohio earthquake of 31 January 1986 : Was it induced?, *Bull. Seismol. Soc. Am.*, *78*, 188-217, 1988.

Nur, A., Dilatancy, pore fluids, and premonitory variation of $ts/tp$ travel times, *Bull. Seismol. Soc. Am.*, *62*, 1217-1222, 1972.

Olami, Z., H. J. S. Feder, and K. Christensen, Self-organized criticality in a continuous, non-conservative cellular automaton modelling of earthquakes, *Phys. Rev. Lett.*, *68*, 1244-1247, 1992.

Ouillon, G., C. Castaing, and D. Sornette, Hierarchical scaling of faulting, *J. Geophys. Res.*, *101*, 5477-5487, 1996.

Pacheco, J.F., C. H. Scholtz, and L. R. Sykes, Changes in frequency-size relationship from small to large earthquake, *Nature*, *355*, 71-73, 1992.

Pearce, R. G., Th relative amplitude method applied to 19 March 1984 Uzbekistan earthquake and its aftershocks, *Phys. Earth Planet Inter.*, *47*, 137-149, 1987.

Pennington, D. W., D. D. Davis, S. M. Carlson, J. Dupree, and T. E. Ewing, The evolution of seismic barriers and asperities caused by the depressuring of fault planes in oil and gas fields of south Texas, *Bull. Seismol. Soc. Am.*, *76*, 939-948, 1986.

Percovic, O., K. Dahmen, and J. P. Sethna, Avalanches, Barkhausen noise and plain old criticality, *Phys. Rev. Lett.*, *75*, 4528-4531, 1995.

Pietronero, L., P. Tartaglia, and Y.C. Zhang, Theoretical studies of self-organized criticality, *Physica A, 173*, 22-44, 1991.

Plichon, J. N., P. Gevin, P. Hoang, P. Londe, and P. Petterville, Sismicité des retenues de grands barrages, in *Proceeding of the 13th Internationnal Large Dams Congress, New Delhi*, pp. 1347-1362, vol. 2 , edited by I.C.O.L.D., Paris, 1979.

Pomeroy, P.W., D.W. Simpson, and M. L. Sbar, Earthquakes triggered by surface quarrying: The Wappingers Falls, New York, sequence of June 1974, *Bull. Seismol. Soc. Am.*, *66*, 685-700, 1976.

Raleigh, C. B., J. H. Healy, and J. D. Bredehoeft, Faulting and crustal stress at Rangely, Colorado, *in Flow and Fracture of Rocks, Geophys. Monogr. Ser.*, vol. *16*, edited by H. C. Heard et al., pp. 275-284, A.G.U., Washington, D.C. 1972.

Roeloffs, E., Fault stability changes induced beneath a reservoir with cyclic variations in water level, *J. Geophys. Res.*, *93*, 2107-2124, 1988.

Roeloffs, E., Hydrologic precursors to earthquakes: A review, *Pure Appl. Geophys.*, *126*, 177-209, 1989.

Romanowicz, B., Strike-slip earthquakes on quasi-vertical transcurrent faults-Inferences for general scaling relations, *Geophys. Res. Lett.*, *19*, 481-484, 1992.

Romanowicz, B., and J. B. Rundle, On scaling relations for large earthquakes, *Bull. Seismol. Soc. Am.*, *83*, 1294-1297, 1993.

Rothé, J.-P, Séismes artificiels (man-made earthquakes), *Tectonophysics*, *9*, 215-238, 1970.
Rydelek, P. A., P. M. Davis, and R. Koyanagi, Tidal triggering of earthquake swarms at Kilauea volcano, Hawaii, *J. Geophys. Res.*, *93*, 4401-4411, 1988.
Saleur H., C. G. Sammis, and D. Sornette, Discrete scale invariance, complex fractal dimensions and log-periodic fluctuations in seismicity, *J. Geophys. Res.*, *101*, 17,661-17,677, 1996.
Saucier, F., E. Humphreys, and R. Weldon, Stress near geometrically complex strike-slip faults: Application to the San Andreas fault at Cajon Pass, southern California, *J. Geophys. Res.*, *97*, 5081-5094, 1992.
Scholz, C. H., Scaling laws for large earthquakes: consequences for physical models, *Bull. Seismol. Soc. Am.*, *72*, 1-14, 1982.
Scholz C. H., Earthquakes and faulting: Self-organized critical phenomena with characteristic dimension., in *Spontaneous Formation of Space-Time Structures and Criticality*, editd by T. Riste and D. Sherrington, Kluwer Acad., pp. 41-56, Norwell, Mass., 1991.
Scholz, C. H., and F. J. Saucier, What do the Cajon Pass stress measurements say about stress on the San-Andreas fault ? Comments on "In situ stress measurements to 3.5 km depth in the Cajon Pass Scientific Research Borehole: Implications for the mechanics of crustal faulting" by Mark D. Zoback and John H. Healy, *J. Geophys. Res.*, *98*, 17867-17869, 1993.
Segall, P., Earthquakes triggered by fluid extraction,. *Geology*, *17*, 942-946, 1989.
Segall, P., Induced stresses due to fluid extractioin from axi-symmetric reservoirs, *Pure Appl. Geophys.*, *139*, 4535-4560, 1992.
Segall, P., J.-R. Grasso, and A. Mossop, Poroelastic stressing and induced seismicity near the Lacq gas field, southwestern France, *J. Geophys. Res., 99*, 15,423-15,438, 1994.
Sethna, J. P., K. Dahmen, S. Kartha, J. A. Krumhansl, B. W. Roberts, and J. D. Shore, Hysteresis and hierarchies-Dynamics of disorder-driven 1st-order phase transitions, *Phys. Rev. Lett.*, *70*, 3347-3350, 1993.
Shimazaki, K., Small and large earthquakes: the effects of the thickness of the seismogenic layer and the free surface, in *Earthquake Source Mechanics, Geophys. Monogr. Ser.*, vol. *37*, edited by S. Das, J. Boatwright, and C. H. Scholz, pp. 209-217, A.G.U., Washington, D.C., 1986.
Sibson, R. H., High-angle reverse faulting in northern New Brunswick, Canada, and its implication for fluid pressure levels,. *J. Struct. Geol.*, *11*, 873-877, 1989.
Silver, P. G., and N. J. Valette-Silver, Detection of hydrothermal precursors to large California earthquakes, *Science*, *257*, 1363-1368, 1992.
Simpson, D. W., Reservoir induced sesismicity, *Eng. Geol.*, *10*, 87-110, 1976.
Simpson, D. W., Triggered earthquakes, *Annu. Rev. Earth Planet. Sci.*, *14*, 21-4, 1986.
Simpson, D. W., and S.K. Negmatullaev, Induced seismicity at Nurek reservoir, Tajikistan,. USSR, *Bull. Seismol. Soc. Am.*, *71*, 1561-1586, 1981.
Simpson D. W. and Leith W., The 1976 and 1984 Gazli, USSR, earthquakes - Were they induced ?, *Bull. Seismol. Soc. Am.*, 75, 1465-1468, 1985.
Simpson D. W.,.W. S. Leith, and C. H. Scholz, Two types of reservoir-induced seismicity, *Bull. Seismol. Soc. Am.*, *78,* 2025-2040, 1988.
Simpson, R. W., S. S. Schulz, L. D. Dietz, and R. O. Budford, The response of creeping parts of the San Andreas fault to earthquakes on nearby faults: Two examples, *Pure Appl. Geophys.*, *126*, 665-685, 1988.
Sornette, A., and D. Sornette, Self-organized criticality and earthquakes, *Europhys. Lett.*, *9*, 197-202, 1989.
Sornette D., Self-organized criticality in plate tectonics, in *The Proceedings of the NATO ASI "Spontaneous Formation of Space-Time Structures and Criticality", Geilo, Norway 2-12 April 1991*, edited by T. Riste and D. Sherrington, Kluwer Acad., pp.57-106, Norwell, Mass., 1991.
Sornette, D., Volcanic tremors given mathematical foundation, *Phys. World,* 23-24, 1992.
Sornette D., Sweeping of an instability : an alternative to self-organized criticality to get powerlaws without parameter tuning, J.Phys.I France 4, 209-221, 1994a.
Sornette, D., Power laws without parameter tuning: An alternative to self-organized criticality, *Phys. Rev. Lett.*, 72, 2306, 1994b.
Sornette, D., and P. Davy, Fault growth model and the universal fault length distribution, *Geophys. Res. Lett.*, 18, 1079-1081, 1991.
Sornette, D., and C. G. Sammis, Complex critical exponents from renormalization group theory of earthquakes : Implications for earthquake predictions, *J. Phys. I.*, *5*, 607-619, 1995.
Sornette, D., and A. Sornette, On scaling for large earthquakes from the perspective of a recent nonlinear diffusion equation linking short-time deformation to long-time tectonics- Comment, *Bull. Seismol. Soc. Am.*, *84*, 1679-1683, 1994.
Sornette, D., and C. Vanneste, Fault self-organization by repeated earthquakes in a quasi-static antiplane crack model, *Nonlinear Processes Geophys.*, *3*, 1-12, 1996.
Sornette, D., and J. Virieux, A theory linking large time tectonics and short time deformations of the lithosphere, *Nature, 357*, 401-403, 1992..
Sornette, D., P. Davy, and A. Sornette, Structuration of the lithosphere in plate tectonics as a self-organized critical phenomenon, *J. Geophys. Res*., *95*, 17353-17361, 1990.
Sornette D., P. Miltenberger, and C. Vanneste, Statistical physics of fault patterns self-organized by repeated earthquakes, *Pure Appl. Geophys.*, *142*, 491-527, 1994.
Sornette, D., P. Miltenberger, and C. Vanneste, Statistical properties of fault patterns self-organized by repeated earthquakes: Synchronization versus self-organized criticality, in the *Proceedings of "Recent Progresses in Statistical Mechanics and Quantum Field Theory"*, edited by P. Bouwknegt et al., pp. 313-332, World Sci., Singapore, 1995.
Sornette, D., L. Knopoff, Y. Y. Kagan, and C. Vanneste, Rank-ordering statistics of extreme events : application to the distribution of large earthquakes, *J. Geophys. Res.*, *101*, 13,883-13,893, 1996.
Spottiswoode, S. M., Source mechanism of mine tremors at Blyvooruitzicht gold mine. in *Proceeding of the 1st International Congress on Rockburst and Seismicity in Mines, Johannesburg, 1982,* edited by N.C. Gay and E.H. Wainwright , pp. 29-37, S. Afr. Inst. of Min. and Metall., Johannesburg, 1984.
Spottiswoode, S. M., and A. McGarr, Source parameters of tremors in a deep-level gold mine, *Bull. Seismol. Soc. Am.*, *65*, 93-112, 1975.
Stein, R. S., G. C. P. King, and J. Lin, Change in failure stress on the southern San Andreas fault system caused by the 1992 *M*=7.4 Landers earthquake, *Science*, *255*, 1687-1690, 1992.
Stein, S., N. H. Sleep, R. J. Geller, S. C. Wang, and G. C. Kroeger, Earthquakes along the passive margin of eastern Canada, *Geophys. Res. Lett.*, *6*, 537-540, 1979.
Volant, P., and J.-R. Grasso, The finite extension of fractal geometry and power law distribution of shallow earthquakes: A geomechanical effect, *J. Geophys. Res.*, *99*, 21,879-21,889, 1994.

Wang, C. Y., and Y. Sun, Oriented microfractures in Cajon Pass drill cores: Stress field near the San Andreas fault, *J. Geophys. Res.*, *95*, 11135-11142, 1990.

Wetmiller, R. J., Earthquakes near Rocky Mountain House, Alberta, and relationship to gas production., *Can. J. Earth Sci.*, *23*, 172-181, 1986.

Wyss, M., and P. Molnar, Efficiency, stress drop, apparent stress, effective stress and frictional stress of Denver, Colorado, earthquakes, *J. Geophys. Res.*, *77*, 1433-1438, 1972.

Yerkes, R. F., W. L. Ellsworth, and J. C. Tinsley,Triggered reverse fault movements and earthquakes due to crustal unloading, northwest Tranverse ranges, California, *Geology*, *11*, 287-291, 1983.

Zhang, Y.-C., Theory of self-organized criticality, *Phys. Rev. Lett.*, *59*, 2125-2128, 1987.

Zhang Y.-C., Scaling theory of self-organized criticality, *Phys. Rev. Lett.*, *63*, 470-473, 1989.

Zoback, M.L., First- and second-order patterns of stress in the lithosphere: The World Stress Map Project, *J. Geophys. Res.*, *97*, 11,703-11,728, 1992.

Zoback, M. D., reply, *J. Geophys. Res.*, *98*, 17871-17873, 1993.

Zoback, M. D., and M. Zoback, State of stress in the conterminous United States., *J. Geophys. Res.*, *85*, 6113-6156, 1980.

Zoback, M. D., and S. Hickman, Insitu study of the physical mechanisms controlling induced seismicity at Monticello Reservoir, South Carolina, *J. Geophys. Res.*, *87*, 6959-6974, 1982.

Zoback, M. D., and B. C. Haimson, *Hydraulic fracturing stress measurements*, 270 pp., U. S. Natl Comm. for Rock Mech., Natl Acad. Press, Washington, D. C., 1983.

Zoback, M. D., and J. H. Healy, Friction, faulting and "in-situ" stress, *Ann. Geophys.*, *2*, 689-698, 1984.

Zoback, M. D., and J. H. Healy, In-situ stress measurements to 3.5 km depth in the Cajon Pass scientific research borehole: Implications for the mechanics of crustal faulting, *J. Geophys. Res.*, *97*, 5039-5057, 1992.

J. R. Grasso, LGIT-IRIGM, Observatoire de Grenoble, BP53X, 38041 Grenoble cedex, France. (e-mail: Jean-Robert.Grasso@obs.ujf-grenoble.fr)

D. Sornette, Laboratoire de Physique de la Matiére Conddensée, CNRS URA 190, Université de Nice-Sophia Antipolis, Parc Valrose, 0608, Nice Cedex 2, France, and Departement of Earth and Space Sciences and Institute of Geophysics and Planetary Physics, University of California 90095-1567, (e-mail: sornette@naxos.unice.fr).

[1]Also at Laboratoire de Physique de la Matière Condensée Université de Nice-Sophia Antipolis, Nice, France

**Figure 1.** Stress field in a grey to black scale, at a given time *t* in the statistically stationary localized regime in a 32 by 32 system with small stress drop parameter and large stress rupture threshold heterogeneity. See *Cowie et al*. [1993] and *Sornette et al.* [1994] for details.

**Figure 2.** Map of the ratio $\sigma/\sigma_c$ of the stress $\sigma$ on each element divided by its rupture threshold $\sigma_c$, for the same system as in Figure 1.

**Figure 3.** Representation, in a grey to black scale, of the active faults at long times i.e. the cumulative slip in the same system as in figures 1 and 2. In this relatively small tectonic plate, only two well-defined faults can be observed.

**Figure 4.** Major seismic events in Gazli area (Uzbekistan), $M \geq 7$ (except for 1978, $M = 5.9$). Black ellipses indicate positions and estimated size of seismic events determined by geodetic data inversion [*Amorese et al.* 1995]. Solidlines represent local directions of the major compressional axis inferred from earthquakes which diverge from the regional NW-SE orientation as determined by *Lukk and Yunga* [1988], *Amorese and Grasso* [1996], and *Bossu and Grasso* [1996].

**Figure 5.** Size of induced earthquakes as a function of water height of the reservoir. Data are for $M \geq 2.5$ reservoir induced seismicity from Table 1 [*Gupta* 1985]. Note that there is no correlation between the size of the triggering process (water height) and the magnitude of the triggered earthquakes.

**Figure 6.** Seismicity induced by reservoir impoundments. The time duration of brittle activation of induced seismicity ranges from a few months after the maximum water level had been reached at Monticello (Figure 6a) to tens of years at Lake Mead (Figure 6b). For the below reservoir induced seismicity, there is no migration, as expected from an elevated isobaric front. The events occur both along and behind the preceeding lower front pressure and are signature of the nonhomogeneous preexisting state of stress and strength. For cases in Figure 6c, 6d, and 6e the latest seismic energy released was within 10 km from the reservoir. The stress change can be read by transforming the water height into pressure (i.e., 100 m ~ 1 Mpa).

**Figure 6a.** Monticello reservoir; number of events per day and cumulative number of events versus lake level (adapted from *Fletcher* [1982]).

**Figure 6b.** Lake Mead, Hoover dam water level (curve) versus monthly frequency of occurrence of earthquakes (vertical bars); adapted from Roeloffs [1988]. The $M\sim5$ largest earthquake occurred in 1939.

**Figure 6c.** Nurek reservoir, Tadjikistan; same as Figure 6b except that the number of $M\geq1.5$ earthquakes is reported monthly. The two $M\sim4.5$ largest earthquakes occurred in 1972. [from *Simpson and Negmatullaev* 1981].

**Figure 6d.** Koyna, India; same as Figure 6b. The $M\sim6.2$ largest earthquake occurred in 1967. The second largest earthquake, $M\sim5.2$, occurred in 1973, when the water level was allowed to increase 1 m beyond the 1967 maximum [from *Simpson , D. W., et al.,*1988].
.
**Figure 6e.** Monteynard reservoir, France; same as Figure 6b except that the number of $M\geq3.5$ earthquakes is reported monthly. The $M\sim5.3$ largest earthquake occurred in 1962. The second largest earthquakes $M\sim4.5$ occurred in 1963 and 1979. The 1979 water level exceeded that in 1963 by <1 m [from *Grasso et al.* 1993].

**Figure 6f.** Acu reservoir Brazil, same as Figure 6b except that the event are reported with a monthly frequency [from *Ferreira et al.* 1995]. In the 1987-1989 interval, the number of events correlates with the water depth, with a 3-month delay. From 1990 on, the correlation between seismicity and water level is no longer clear.

**Figure 7.** Seismicity activated by subsurface reservoir depletions . (a) Strachan field, Canada; number of earthquakes recorded per year and decline in average gas reservoir pressure, adapted from *Segall* [1989]. The $M\sim4$ largest earthquake occurred in 1974. Data from *Wetmiller* [1986]. (b) Lacq field, France; number of $M\geq3.0$ earthquakes recorded per year and decline in average gas reservoir pressure. The $M\sim4.2$ largest earthquake occurred in 1981; adapted from *Grasso and Wittlinger* [1990]. (c) Ekofisk field, Norway; number of $M\geq2.5$ earthquakes recorded per year and decline in average gas reservoir pressure. The $M\sim3.8$ largest earthquake occurred in 1988. Pressure data from Phillips Oil (personal. communication, 1992). (d) Assen field, northern Holland; number of $M\geq2.5$ earthquakes recorded per year and decline in average gas reservoir pressure. The $M\sim2.8$ largest earthquake occurred in 1986. Data for earthquakes from *Haak* [1991]. (e) Fashing field, southern Texas; number of $M\geq2$ earthquakes recorded per year and decline in average gas reservoir pressure. The $M\sim3.4$ largest earthquake occurred in 1983. Data from *Pennington et al* [1986]. f) Imogene field, Southern Texas; number of $m\geq2$ earthquakes recorded per year and decline in average gas reservoir pressure. The $M\sim3.9$ largest earthquake occurred in 1984. Data from Pennington et al (1986). The induced stress (Figure 7a-7f) at a time $t$ is roughly related to the gas reservoir pressure as follows (see text for details): $\Delta\sigma\sim3 \times 10^{-3} (P_{init}-P_t)$. Note that if the seismicity is triggered at the value $\Delta\sigma\sim1$ MPa of induced crustal stress change for all the case studies, it is sustained for $P_{init}-P_t \leq 10$ MPa, i.e., $\Delta\sigma\leq0.1$ MPa.

**Figure 7.** (continued)

**Figure 8.** Example of isolated induced seismicity. The volcanic seismicity (1988-1992) exibits both (a) a power law distribution of earthquake sizes and (b) a random temporal relationship with the Earth tidal cycle (adapted from *Grasso and Bachelery* [1995]).

**Figure 9.** Temporal pattern of mining induced seismicity. For South Africa gold mines, the seismic events disappear during weekends and holidays. Without any strong forcing (the working day excavation rate) of the system, earthquake faulting disappears [from *Cook*, 1976].

**Figure 10a.** Example of power law size distribution relation for induced earthquakes. Mining induced seismicity, South Africa gold mines. Seismic moment as a function of corner frequency for mine earthquakes in South Africa. Data from *Spottiswoode and McGarr* [1975], Circles; *McGarr et al.* [1981], diamonds; and *Spottiswoode* [1984], squares. Adapted from *Spottiswoode* [1984].

**Figure 10b.** Fluid extraction induced seismicity, Lacq gas field, France. Seismic moment as a function of source radii. The source radii are calculated using a rupture velocity of $0.9\beta$, $R=0.32\beta /f_0$. From *Feignier and Grasso* [1992].

**Figure 10c.** Reservoir induced seismicity at Oroville reservoir, California. Peak velocity parameter ($Rv$) as a function of seismic moment. Note that $Rv$ scales with the source radii for constant stress drop scaling, i.e., $Rv \sim m_0^{1/3}$. Adapted from *McGarr* [1986].

**Figure 11.** Time duration of seismicity forcing induced by subsurface mining. Example of a South Africa gold mine and constant break slope in self-similarity (adapted from *McGarr* [1976]).

**Figure 12.** Finite size effects in mining induced seismicity. (a) Rolloff in the frequency-magnitude distribution reported in coal mining induced seismicity;data from *Kusznir and Al-Saigh* [1984]. (b) Rolloff in the frequency-

magnitude distribution reported in gold mine induced seismicity from underground seismic observations (data from *Deliac and Gay* [1984]).

**Figure 13a.** Finite size effects in reservoir induced seismicity.
   Rolloff in the frequency-magnitude distribution reported for Monteynard Reservoir, France. Data (squares) are from the seismic station that operated within the dam during the December 1963 to December 1967 period. Note that both the $M=4.3$ shock that occurred in August 1966 and the sequence of three $m \geq 4$ earthquakes that occurred in the April 25-27, 1963, period (triangles) are shifted relatively to the number the small shocks ($3.1 \geq M \geq 1$). All the earthquakes are within 7 km from the reservoir (data from *Plichon et al.* [1979]). Note that the observed frequency-size relationship of Monteynard earthquakes mimics the laboratory observations (Figure 13b) of *Krantz et al.* [1990]. Accordingly, the low number of large earthquakes in the December 1963 to december 1967 period could correspond to fractures within a nonfluid-pressured area whereas the $M \geq 4$ shocks could be a direct effect of pore pressure-induced seismicity.

**Figure 13b.** Number of events having an average normalized peak amplitude greater than *A* (in units of mV/mm) before and after borehole pressurization in rock sample experiments [from *Krantz et al.*1990].

**Figure 13c.** Geological cross section of the local structure where brittle calcaerous beds and ductile marly beds, *Vm* and *TN* levels, respectively) are reported. The main shocks and their focal mechanisms are plotted. Adapted from *Grasso et al.* [1993].

**Figure 14a** Finite size effects of local activation of earthquakes. Frequency-size relationship for induced earthquakes in the vicinity of Lacq gas field.

**Figure 14b.** Fractal dimension of the same data set (adapted from *Volant and Grasso* [1994]).

**Figure 14c.** Geomechanical cross section of the local structure where deep brittle calcareous beds ($E$=50 GPa, $\sigma_c$= 100 MPa) and ductile marly beds ($E$= 30 GPa, $\sigma_c$= 70 MPa) are reported. Mechanical properties are from *Fabre et al.* [1992]. Note the correlation between the break slope point in both *b* value and fractal dimension at a common 500 m distances ($M$~3) that is roughly the thickness of seismogenic beds.

**Table 1b.** Hydromechanical Setting of Earthquakes Triggered by Unloading of the Upper Crust Caused by Hydrocarbon Extraction.

| Area or Field | Reservoir Depth, km | Reservoir Area, km$^2$ | 0nset of Production - Earthquakes | Initial Pressure, MPa | Mass Removal, $10^{11}$ kg | Earthquake size M - $m_o$, mN | Induced Stress (Unloading), MPa | Vertical Stress Change (Unloading), MPa |
|---|---|---|---|---|---|---|---|---|
| Caviaga field Po Valley, Italy | 1.5 | - | 19?? - 1951 | 13 | - | 5.5 - $10^{17}$ | - | - |
| Coalinga field California, | 2 | 110 | 1940 - 1983 | 25 | 2.7 | 6.5 - $4 \times 10^{18}$ | 0.01 | 0.024 |
| Kettleman field California, | 1.5 | 50 | 19?? - 1985 | - | 1.2 | 6.1 - $2 \times 10^{18}$ | 0.01 | 0.024 |
| Montebello field California, | 1.5 | 6 | 1924 - 1987 | - | 1.3 | 5.9 - $1 \times 10^{18}$ | 0.01 | 0.21 |
| Gazli field Uzbekistan | 1.5 | 300 | 1962 - 1976 | 10 | 1.9 | 7.0 - $2 \times 10^{19}$ | 0.04 | 0.006 |
| Gazli field Uzbekistan | 1.5 | 300 | 1976 - 1984 | 3 | 1.8 | 7.0 - $2 \times 10^{19}$ | 0.05 | 0.006 |

The year when extraction began and the date of the first induced event, mass deficit and the size of the largest induced event, are given. Assuming that mass withdrawal at these sites is related to the major earthquake, we estimate the induced stress (augmenting failure) by multipliying the earthquake stress drop by the ratio of stiffness of the loading system (the vertical force change) to the stiffness of circular crack model according to *McGarr* [1991]. Note that the initial pressures of the reservoirs induce poroelastic stress changes that are 1 order of magnitude less than the stress change induced by mass removal. For the Coalinga field, the poroelastic stress change and the stress change due to unloading have a value of the same magnitude order at the hypocentral depth of the mainshock [*McGarr*, 1991].

**Table 1a.** Mechanical Constants used to Estimated the Poroelastic Stress Change that Triggered Seismicity in the Neighborhood of Hydrocarbon Extraction.

| Area or Field (Unloading) | Reservoir Characteristic | | | | Subsidence Maximum, m | Pressure, MPa | | Onset of Production-eqs, years | Maximum Magnitude $M$ | Induced Stress (Poroelastic), MPa | Mass Change, $10^{11}$ kg | Reservoir Area, km$^2$ | Stress Change (Vertical Unloading) MPa |
|---|---|---|---|---|---|---|---|---|---|---|---|---|---|
| | $D$, km | $T$, km | $\mu$, GPa | $\nu$ | | Initial | Change | | | | | | |
| Strachan field Alberta, Canada | 3-5 | 0.1 | - | - | - | 50 | -25 | 1971-1976 | 3.4 | ≤0.7[a] | 0.18 | 25 | 0.007 |
| Fashing field Texas | 3.4 | 0.03 | - | - | - | 35 | -23 | 1958-1974 | 3.4 | ≤0.7[a] | <0.1 | 10 | <0.01 |
| Imogene field Texas | 2.4 | 0.03 | - | - | - | 25 | -13 | 1938-1973 | 3.9 | ≤0.4[a] | <0.1 | 20 | <0.005 |
| Lacq field Aquitaine, France | 3-6 | 0.25 | 23 | 0.25 | .005 | 66 | -30 | 1959-1969 | 4.2 | ≤0.5[b] | 2.0 | 150 | 0.01 |
| Grozny field Tchetcheny, Caucasus | 4-5 | 0.67 | - | - | - | 69 | -25 | 1964-1971 | 4.1 | ≤0.7[a] | 1.5 | 70 | 0.02 |
| Assen field Netherlands | 3.0 | 0.1 | 10 | 0.25 | - | 30 | -25 | 1972-1986 | 2.8 | ≤0.7[b] | <0.1 | 50 | 0.002 |
| Groningen field Netherland | 2.9 | 0.2 | 10 | 0.26 | 0.16 | 35 | -30 | 1964-1991 | 2.5 | ≤0.4[b] | 9.0 | 900 | 0.01 |
| Ekofisk field offshore, Norway | 3-5 | 0.2 | 0.07 | 0.30 | 3.50 | 48 | -24 | 1973-1982 | 3.4 | ≤0.4[b] | 0.2 | 130 | 0.002 |
| Dan offshore, Danemark | 1.8 | 0.15 | 0.30 | - | - | 26 | -18 | 1972-1985 | 4.0 | ≤0.45[b] | <0.1 | 24 | 0.004 |

The reservoir characteristics used are depth ($D$), thickness ($T$), shear modulus ($\mu$) and Poisson's ratio ($\nu$). The maximum depth of the subsidence bowl at the onset of the seismicity, initial reservoir pressure and pressure drop at the onset of seismicity, the onset of production, the onset of seismicity, the maximun magnitude reported in the field, and the maximum induced stress at the onset of seismicity, estimated using poroelastic stressing are given.

[a] Subsidence value or rock properties are not reported, we estimated $\Delta\sigma_{max}$ using equation (2); $\Delta\sigma_{max} \approx [(1-2\nu)/2\pi(1-\nu)] \Delta p \, F_{max}(a/D)$, where $\Delta p$, $F_{max}(a/D)$ are the reservoir pressure drop and a dimensionless function of position and reservoir geometry, respectively, $a$ is the reservoir radius, $D$ is the reservoir depth (see text for details).

[b] The subsidence and shear modulus of the rocks involved are known and we used equation (1) in order to estimate $\Delta\sigma_{max} \approx [4\mu/(1-\nu)\pi D] \Delta H_{max}$, where $\mu$, $\nu$, $D$, and $H_{max}$, are the shear modulus, the Poisson coefficient of the rock matrix, the depth of reservoir, and the maximum depth of the subsidence bowl, respectively.

Note that for the all reservoirs, the poroelastic stress change is 1 order of magnitude larger than the vertical stress change induced by mass discharge, $-\Delta mg/A$; $A$ being the effective reservoir area. Modified from *Grasso* [1993].

**Table 2.** *M*>5 Induced Earthquakes.

| Dam / Reservoir or Hydrocarbon Field | $\Delta P$, MPa | $\Delta M$, $10^{11}$ kg | Maximum Magnitude | Regional Seismicity | Orientation of $\sigma_1$ From Induced Earthquakes | Tectonic Setting |
|---|---|---|---|---|---|---|
| Gazli field, EIS Uzbekistan | ≤0. | -2. | 7.3 | low | horizontal | midplate |
| Koyna, RIS India | +1. | +27. | 6.5 | low | horizontal | midplate |
| Coalinga field, EIS USA | ≤0. | -2.7 | 6.5 | high | horizontal | plate boundary |
| Kremasta[a], RIS Greece | +1.6 | +47. | 6.3 | high | vertical | back arc extension |
| Hsinfengkiang, RIS China | +1. | +1.0x$10^2$ | 6.1 | low | horizontal | midplate |
| Kettleman field, EIS USA | ≤0. | -1.2 | 6.1 | high | horizontal | plate boundary |
| Montebello field, EIS USA | ≤0. | -1.3 | 5.9 | high | horizontal | plate boundary |
| Oroville, RIS USA | +2.4 | +43 | 5.9 | low | vertical | Sierra Nevada foothills |
| Kariba, RIS Zambia/Zimbabwee | +1.3 | +1.7x$10^3$ | 5.8 | low | vertical | midplate |
| Marathon[a], RIS Greece | +0.7 | +0.4 | 5.7 | high | n.a | plate boundary |
| Aswan[a], RIS Egypt | +1.1 | +1.6x$10^3$ | 5.5 | low | vertical | midplate |
| Eucumbene, RIS Australia | +1.1 | +47 | 5.5 | low | n.a | midplate |
| Hoover, RIS USA | +2.2 | +3.7x$10^2$ | 5.5 | low | vertical | Colorado plateau |
| Denver, IIS USA | +3.5[b] | +$10^{-3}$ | 5.5 | low | vertical | Colorado plateau |
| Caviaga, EIS Italy | ≤0. | -?? | 5.5 | low | horizontal | midplate |
| Lake County, IIS USA | +0.5[c] | ~0. | 5.3 | low | horizontal | midplate |
| Monteynard, RIS France | +1.3 | +2.7 | 5.3 | low | vertical | Alps foothills |
| El Reno, EIS USA | ≤0. | -?? | 5.2 | low | horizontal | midplate |
| Snipe Lake, EIS Canada | ≤0. | -?? | 5.1 | low | horizontal | midplate |

n.a., noavailable information. RIS, EIS, and IIS arereservoir-, extraction-, and injection- induced seismicity, respectively. See references for RIS, EIS, and IIS within *Gupta* [1985], *Grasso* [1993], and *Nicholson and Wesson* [1993] respectively.

[a] Disagree with decision diagramm for Corps of Engineers study of RIS [*Meade*, 1991].
[b] Modeled fluid pressure increase [*Hsieh and Bredehoeft,* 1981[.
[c] Modeled fluid pressure increase [*Nicholson et al*., 1988[.